\documentclass[useAMS,usenatbib]{mnras}
\usepackage{amsmath,amsfonts,graphicx,txfonts,fleqn,color,tabularx,verbatim,fixltx2e}
\newcommand {\B}[1] {\boldsymbol{#1}}

\newcommand  *{\D}[1] {\dot{\B{#1}}}
\newcommand  *{\DD}[1] {\ddot{\B{#1}}}

\title[There might be giants]{There might be giants: unseen Jupiter-mass planets as sculptors of tightly-packed planetary systems}
\author[Hands \& Alexander]{T.O.Hands\thanks{email:
tom.hands@le.ac.uk} and R.D.Alexander 
\\Department of Physics \& Astronomy, University of Leicester, University Road, Leicester, LE1 7RH, UK}
\begin{document}
\voffset=-0.25in

\pagerange{\pageref{firstpage}--\pageref{lastpage}} \pubyear{2014}

\maketitle

\label{firstpage}

\begin{abstract}
The limited completeness of the {\it Kepler} sample for planets with orbital
periods $\gtrsim$ 1 yr leaves open the possibility that 
exoplanetary systems may host undetected giant planets. Should
such planets exist, their dynamical interactions with the inner planets
may prove vital in sculpting the final orbital configurations of these
systems. Using an $N$-body code with additional forces to emulate the
effects of a protoplanetary disc, we perform simulations of the
assembly of compact systems of super-Earth-mass planets with unseen giant
companions. The simulated systems are analogous to
Kepler-11 or Kepler-32 in that they contain 4 or 5 inner super-Earths, but our systems also contain longer-period giant
companions which are unlikely to have been detected by {\it Kepler}. We find that giant companions tend to break widely-spaced, first-order mean-motion resonances, allowing the inner planets to migrate into tighter resonances. This leads to more compact architectures and increases the occurrence rate of Laplace resonant chains.
\end{abstract}


\section{Introduction}
The recent explosion of results in the field of extra-solar planet detection
has revealed several new and distinct populations of planets that are of great
interest from a dynamical perspective. In particular, the {\it Kepler} mission
has discovered a multitude of compact systems, each consisting of 5 or 6
planets in the super-Earth to Neptune mass regime and all orbiting within
$1\mathrm{AU}$ of their host star \citep[see e.g.,][]{Lissauer2011,Swift2013,Quintana2014}. This class of planets appears to be very common. Both radial velocity surveys and the data from the {\it Kepler} mission agree in suggesting that $\gtrsim$50\% of stars host at least one close-in super-Earth \citep{Chiang2013}. Multiplicity is also very common. The sixth {\it Kepler} data release shows that 1640 (39.3\%) of a total of 4175 Kepler Objects of Interest are in multiple-planet systems, with 656 (20.6\%) of a total of 3191 candidate systems containing multiple planets \citep{Mullally2015}. These figures are an increase from the 38.4\% and 19.9\% reported respectively for these statistics in the previous data release \citep{Burke2014}, and include perhaps the most extreme example of a compact system yet: Kepler-90. This system contains 7 planets, two of which are gas giants, the outermost of which orbits its host star at 1.01AU and has a radius roughly equivalent to that of Jupiter \citep{Cabrera2014,Schmitt2014}. 

Radial velocity surveys have revealed a
large population of Jupiter-mass objects orbiting their stellar hosts exterior
to $1\mathrm{AU}$ \citep[see e.g.,][]{Marmier2013}. The {\it Keck} survey suggests that between 17--20\% of Sun-like stars could host gas giant planets within $20\mathrm{AU}$  \citep{Cumming2008}, whilst {\it HARPS} finds that 14\% of such stars host a gas giant in an orbit of 10 years or shorter \citep{Mayor2011}. In spite of the relatively high incidence of both close-in super-Earths and gas giants, Kepler-90 represents the only known example of a system which contains both of
these types of planets. This is likely to be a result of the limited sensitivity of each detection method, with {\it Kepler} having limited completeness exterior to $1AU$ and radial velocity surveys being both insensitive to lower-mass planets and unable to perform follow up on dim, distant {\it Kepler} targets  \citep[see][for a recent review]{Fischer2014}. It is unlikely that the two populations are mutually exclusive, and future observational campaigns may shed light on the overlap between them.

In our previous work \citep{Hands2014} we considered the possibility of
assembling compact systems of super-Earths via disc-driven (Type I)
migration. We found that this method can reliably produce systems analogous to
Kepler-11 or Kepler-32, albeit with a greater occurrence of mean-motion
resonances than in the observed systems. In this paper we explore the possibility
that some compact systems may contain additional, undetected Jupiter-mass
companions orbiting exterior to the known planets. We run a suite of numerical
simulations of the assembly of such systems. The premise is that these planets form much further out in the disc and then migrate inwards in the Type I
regime as a result of their gravitational interaction with the gas disc. In each simulation we allow one of the outer embryos to undergo runaway growth, rapidly accreting gas from the disc and growing exponentially in mass to become a gas giant. We follow the evolution of these systems numerically, and investigate the effects of outer giant planets on the orbital architectures of the observable (inner) planets.

\begin{figure*}
\begin{center}
\includegraphics[width=0.74\linewidth]{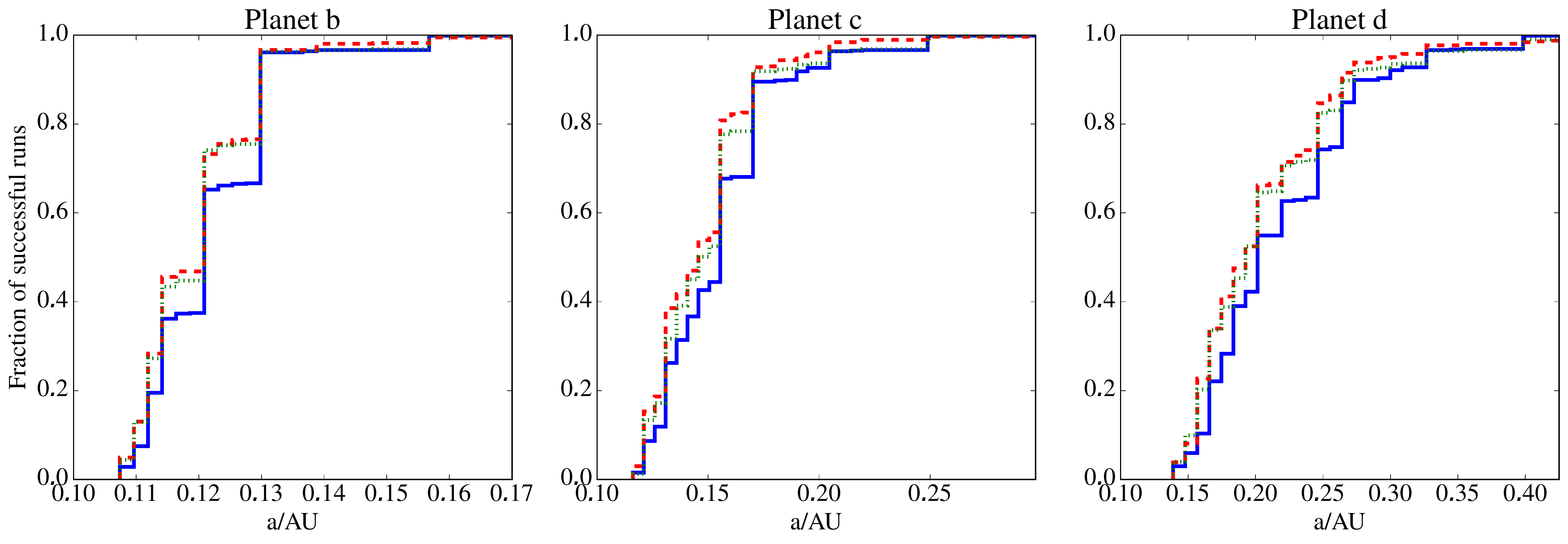}
\includegraphics[width=0.99\linewidth]{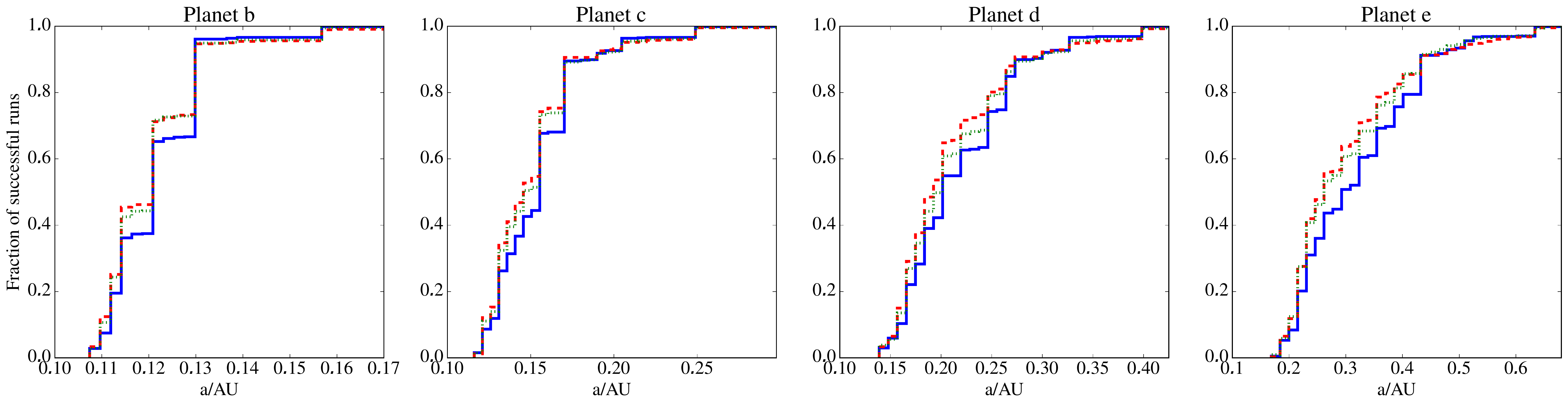}
\caption{Comparison between distributions of final semi-major axes of interior, super-Earth mass planets in the control case with no giant (blue/solid), the case with planet e (top row) or f (bottom row) planet becoming a $1\mathrm{M_{jup}}$ giant at 1AU (green/dotted) and the case with planet e/f becoming a $3\mathrm{M_{jup}}$ giant at 1AU (red/dashed). The linear growth case is not plotted. The interaction with the giant generally allows the lower-mass planets to occupy tighter orbits by breaking widely-spaced resonances between them. \label{fig:distdiff} }
  \end{center}
\end{figure*}

\begin{figure*}
\begin{center}
\includegraphics[width=0.33\linewidth]{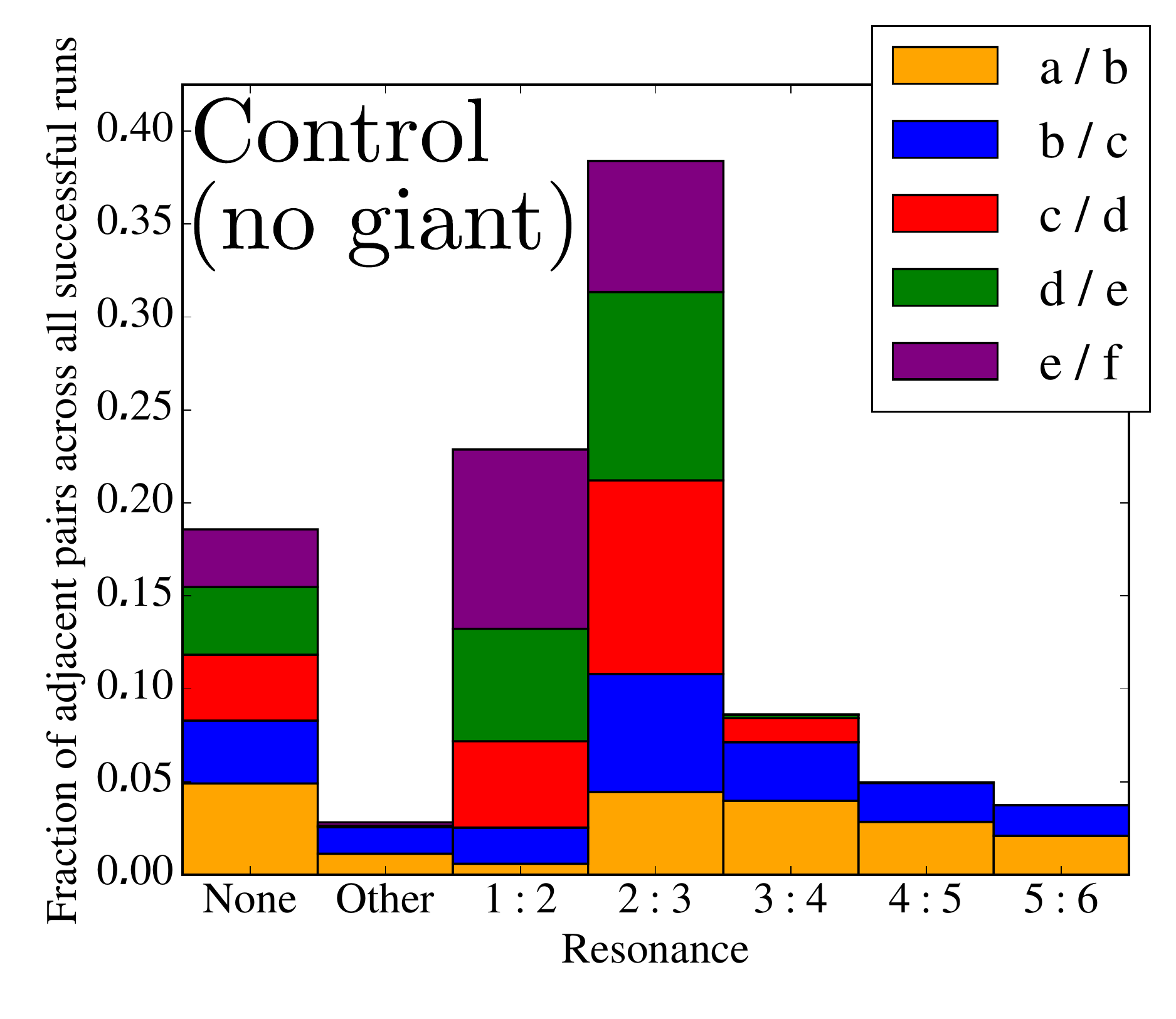}
\includegraphics[width=0.33\linewidth]{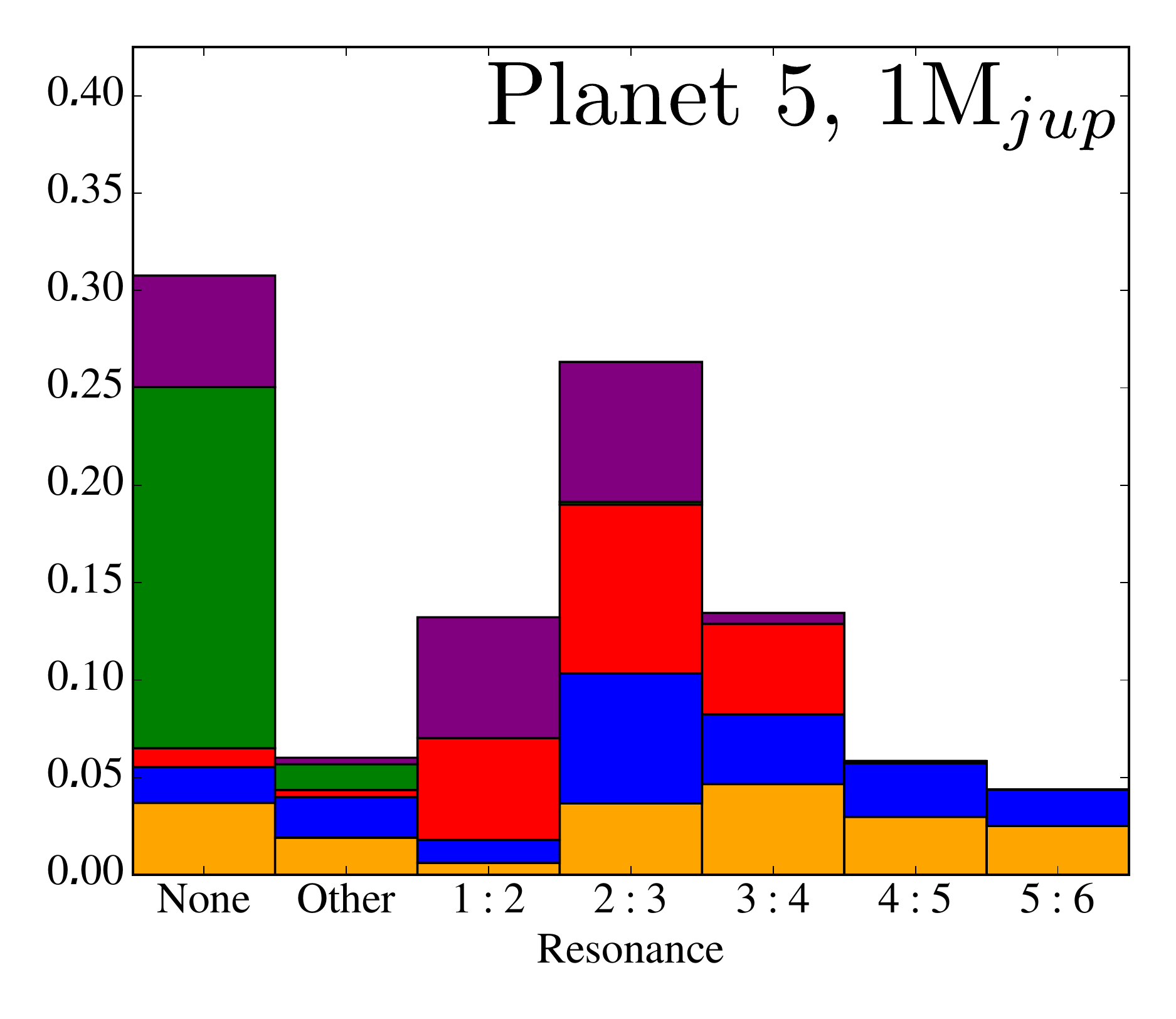}
\includegraphics[width=0.33\linewidth]{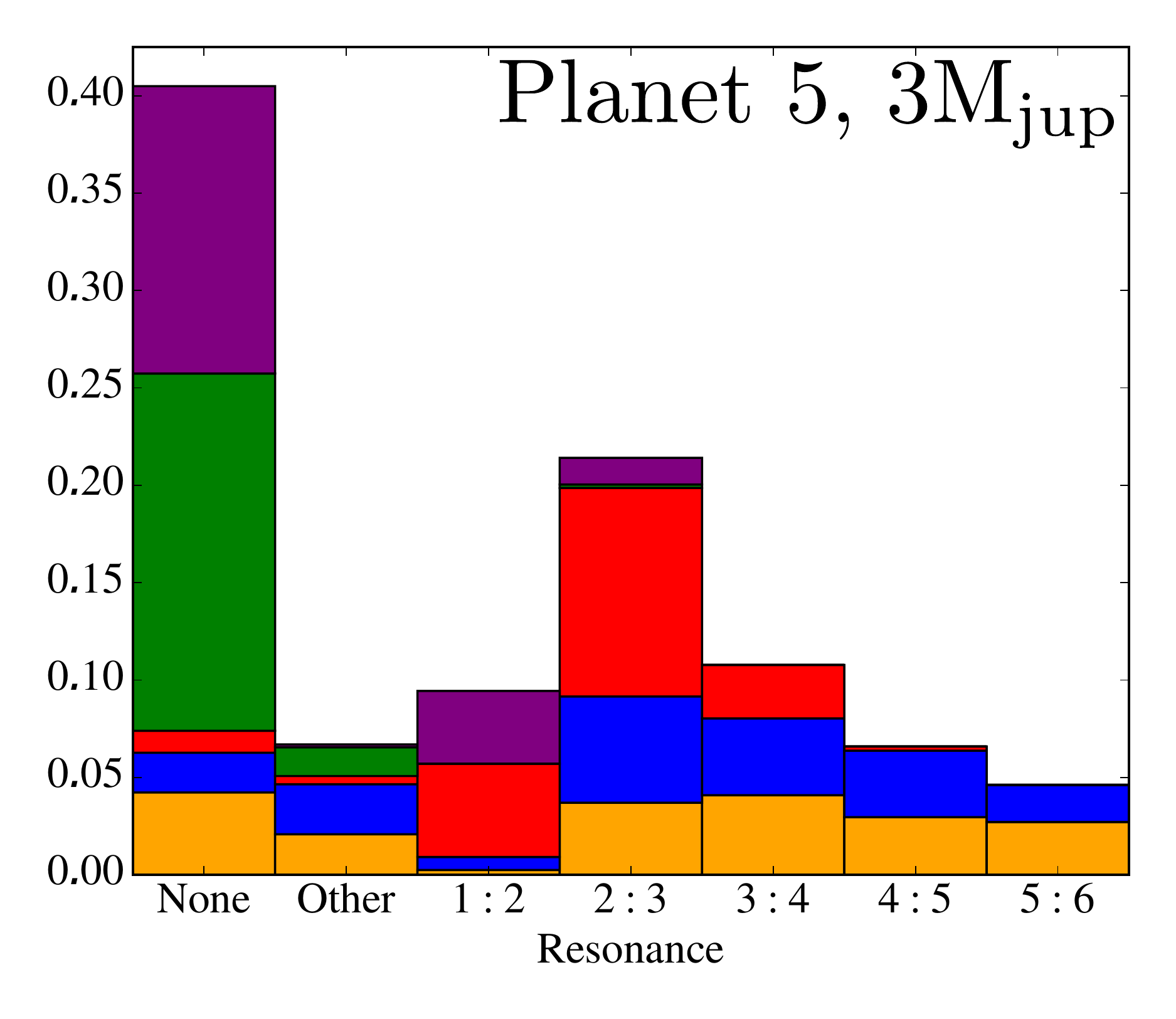}
\includegraphics[width=0.33\linewidth]{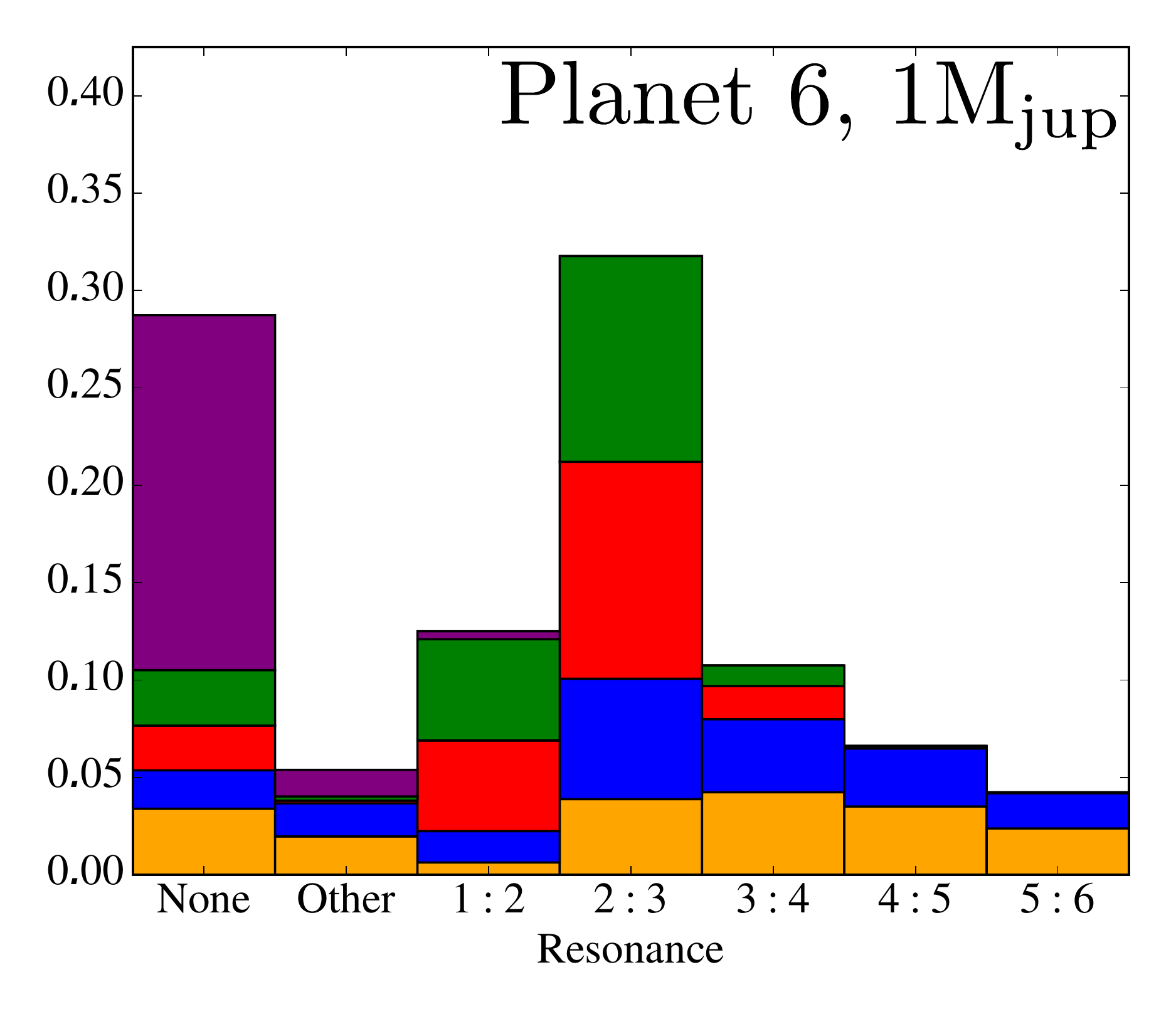}
\includegraphics[width=0.33\linewidth]{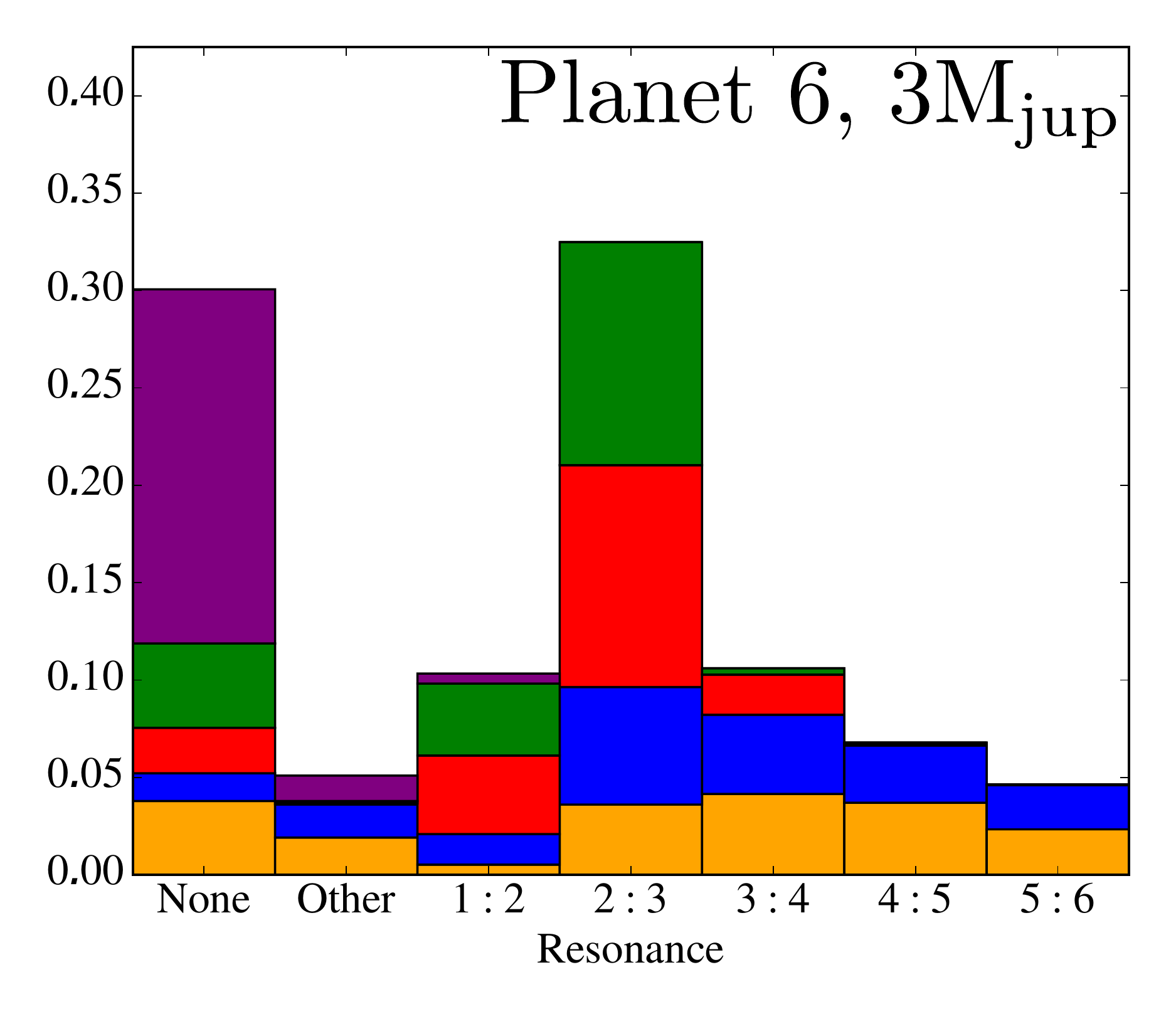}
\includegraphics[width=0.33\linewidth]{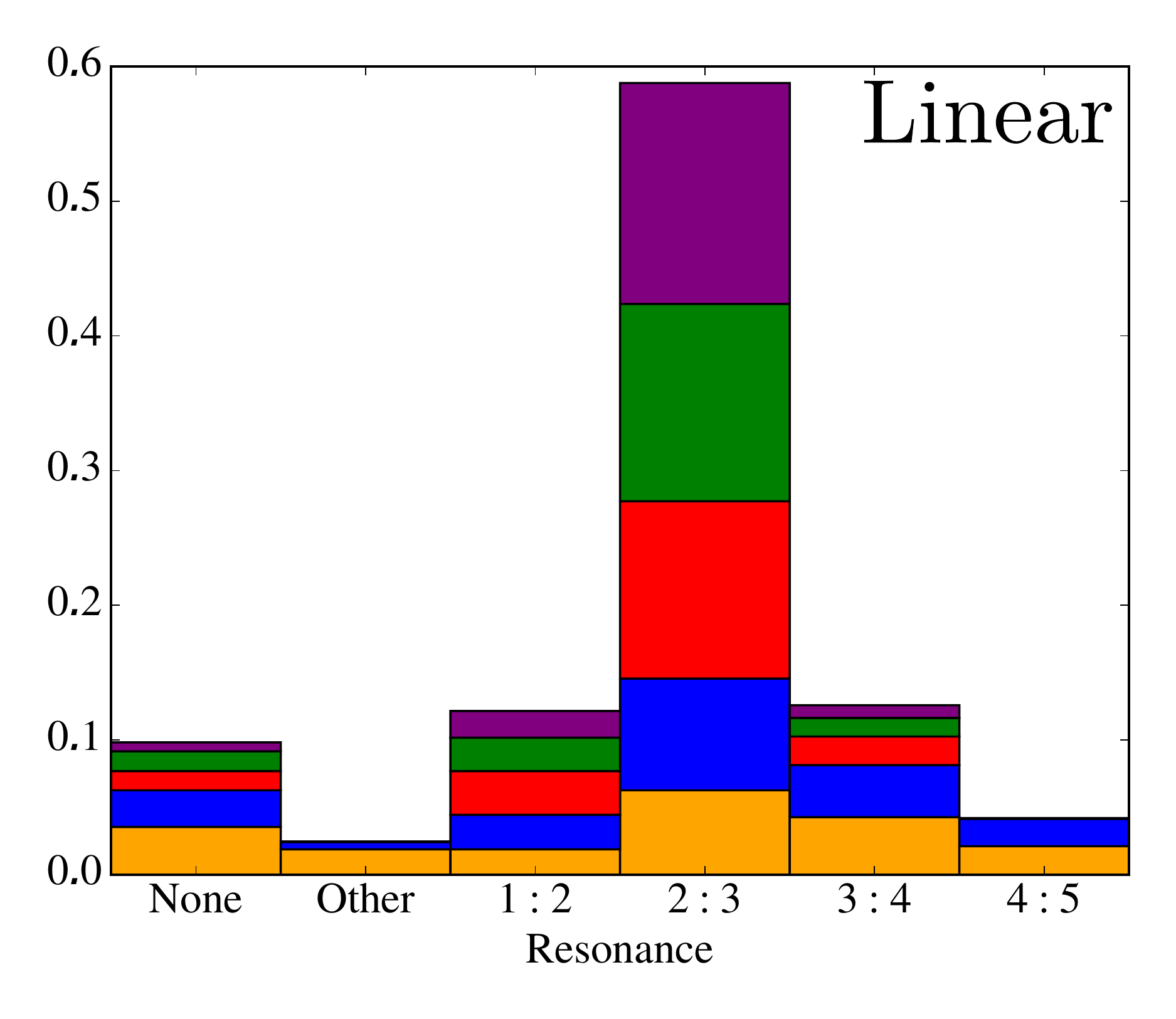}
\caption{ Fractions of adjacent pairs of planets that are in each mean-motion resonance at the end of our successful runs.  Planets are labelled a--f in order of increasing semi-major axis. The "Other" bar is a sum over all resonances that contain less than 3\% of the total number of pairs. The numbers in the key show which two adjacent planets are represented by each colour. The simulations represented in each plot are (clockwise from top-left) the control, planet e becoming a $1\mathrm{M_{jup}}$ giant, planet e becoming a $3\mathrm{M_{jup}}$ giant, the linear growth model, planet f becoming a $3\mathrm{M_{jup}}$ giant and planet f becoming a $1\mathrm{M_{jup}}$ giant. In comparison to the control case, the cases with a giant companion exhibit fewer widely-spaced mean motion resonances, and more tight ones. For instance, note the relative increase in the occurrence of the $3:4$ resonance compared to the $2:3$ resonance between planets a and b between the control case and the other cases. Note also that the scale on the y-axis of the bottom-right plot is different to the others.\label{fig:resdist} }
  \end{center}
\end{figure*}

\section{Numerical method}
We follow the method of \cite{Hands2014}, simulating the migration of multiple
super-Earth mass planets using an $N$-body integrator with imposed
migration forces. This method is based on that of \cite{Rein2009}, and imposes an exponential decay of both semi-major axis and eccentricity on each planet, while simultaneously adding a stochastic forcing component in the $r$ and $\theta$ directions to simulate the effect of disc turbulence. These forces are controlled by 3 free parameters: $\tau$, which sets the migration time-scale of a planet; $K$ which sets the ratio between the eccentricity-damping time-scale and the migration time-scale; and $\beta$, which sets the RMS strength of the stochastic forces relative to the gravitational force exerted on each planet by the host star. Note that $\tau$ also scales inversely proportional to the mass of each planet, meaning that higher-mass planets migrate faster, as is expected of the Type I regime. For the purposes of this study we vary $\tau$ between $10^{3.5}$ and $10^{5.5}\mathrm{yr}$, $K$ between $10^{1.5}$ and $10^{2.5}$, and $\beta$ between $10^{-6}$ and $10^{-8}$. A discussion of the physicial significance of these parameter values can be found in \cite{Hands2014}. Note that the overall range of the parameter space is reduced in this work, since in our previous work we showed that relatively high levels of stochastic forcing and relatively low values of eccentricity damping led to almost all compact systems being disrupted during their assembly. 

\subsection{Runaway gas accretion}
In addition to parametrized disc forces, we include a simple prescription for the runaway gas accretion, to simulate the rapid growth of a protoplanet from a large super-Earth up to a Jupiter-mass giant. This process is an important part of the core accretion theory of planet formation \citep[see e.g.][for a recent review]{Helled2014}, and begins after a protoplanet has been accreting gas slowly on to its envelope for some time (typically $\sim$Myr). Once the envelope mass becomes comparable to the core mass the envelope contracts rapidly and matter is accreted on to the planet as fast as local disc conditions permit, until the planet becomes massive enough to open a gap in the disc \citep{Pollack1996}.

In our model, the planet initially maintains a constant mass $M_0$ as it migrates inwards. When it reaches some arbitrary distance from the star, the planet mass grows as
\begin{equation}\label{growth}
M = M_0 \mathrm{exp}\bigg{(}\frac{t}{\tau_g}\bigg{)},
\end{equation}
where $M$ is the mass of the planet at time $t$ after the growth process begins, and $\tau_g$ is a characteristic growth timescale. This timescale naturally varies with the disc parameters in the vicinity of the planet, with \cite{Dangelo2008} finding from two- and three-dimensional hydrodynamical simulations that this process takes several hundreds of orbits. We hence set $\tau_g$  to $10^3 \mathrm{yr}$. The migration and eccentricity damping forces and stochastic forces for the planet in question are switched off at the point when runaway growth begins. While in reality such a planet would likely continue to migrate inwards in the gap-opening Type II regime, the rate of such migration is sufficiently slow compared to the Type I regime that the movement would be negligible compared to the other super-Earths in the system \citep[e.g.,][]{Baruteau2014}.

Since the migration force is turned off once runaway growth begins, the point at which this process is started essentially sets the radius at which all giant planets in our models will orbit. We select 1 AU as this distance, giving the giants a period of about a year. This radius was selected due to it being the minimum distance at which we would expect {\it Kepler} to not see the majority of planets, since at this radius even a small inclination will lead to a planet not transiting. This means that all of the giant planets in our results orbit in a small annulus around 1AU, with minor deviations being caused by dynamical interactions with the other planets during the growth process.

\subsection{Linear mass accretion}
As an alternative to the model in which one planet grows exponentially, we test a scenario in which the outermost 4 planets grow from low initial masses - approximately equivalent to the masses of the innermost 2 planets - to their larger, super-Earth masses as they migrate inwards. This ensures that any behaviour seen in our simulations with exponential gas accretion are truly down to the dynamical effect of the giant, and not simply an artefact of our growth prescription. In this case, the mass of each planet increases linearly at a rate

\begin{equation}
\frac{dM}{dt} = \frac{M_{final} - M_{initial}}{5 \tau},
\end{equation}
where $\tau$ is the global migration timescale for that simulation and is itself unrelated to the mass of the planet.

\subsection{Initial conditions} \label{sec:init}
For the purposes of these simulations, we use a model planetary system designed to be analagous to compact systems such as Kepler-11 or Kepler-32. This system consists of 6 planets that are initially in the super-Earth to Neptune mass regime. The masses are based loosely on those of Kepler-11, and assigned such that the planets are mass-ordered from lowest- to highest-mass with increasing distance from the star, hence emulating the configuration of other compact systems. From inner- to outer-most, the planets in this system have initial masses of 1.9, 2.9, 7.3, 8.0, 8.7 and 9.3 $\mathrm{M_\oplus}$ respectively, and we label them a--f in order of increasing initial mass/semi-major axis. The jump in mass between the second and third planets is similar to that observed in Kepler-11 \citep{Lissauer2013}. Radii are then assigned according to the masses of the planets, using the power-law
${M_p}/{M_{\oplus}}= ({r_p}/{r_{\oplus}})^{2.06}$  \citep{Lissauer2011a},
although planetary radii are used only for detecting collisions.

The inital positions of the planets are picked in a similar way to
\cite{Hands2014}, with the innermost planet being placed just exterior to the
expected snow-line (at 1.5AU) and each subsequent planet being separated from
its neighbours by an oligarchic spacing argument
\citep[e.g.,][]{Kokubo1998}. Pairs of adjacent planets are
separated by a random number of mutual Hill radii ($r_h$), picked from a
normal distribution. Here we use a mean of 28$r_h$ with a standard-deviation
of 5$r_h$, selected such that adjacent planets are in general initially
situated exterior to the 2:1 resonance. Initial phases for each planet are
selected at random from a uniform distribution, and all simulations are co-planar.

\subsection{Models}
We run 6 distinct sets of models in total. In the first 4 we vary which of the outer two planets
undergoes exponential growth into a giant (planet e or f), and the final mass of the giant (1 or
3 $\mathrm{M_{jup}}$). In the fifth set of models, we grow the outermost 4 planets from 2.93, 3.00, 3.06 and 3.12 $\mathrm{M_\oplus}$ respectively to their final masses (7.3, 8.0, 8.7 and 9.3$\mathrm{M_\oplus}$ respectively) using the linear growth model. The final set of models
is a control, in which none of the planets grow into a giant. Each set consists of 1000 individual models, with initial positions and phases of the planets being varied randomly between each. Values for each of the three forcing parameters are also picked randomly and uniformly in log space from the prescribed range.

Each model runs until the innermost planet (planet a) in the system has reached 0.1
$\mathrm{AU}$, chosen for its similarity to the semi-major axis of the
innermost planet in Kepler-11. Models in which planets collide or a planet
is ejected from the system are discarded, as are the small minority of models in which two planets switch positions.

\subsection{Analysis}
For each of the sets of models in which a giant planet was formed, we compare
the distribution of the final position of each individual planet to the control case using a Kolmogorov--Smirnov (K-S) test. For each planet in each set of models this yields the probability that the presence of the giant significantly affects the final location of the planet. The results from this analysis are shown in table \ref{tab:kstable}.

We also apply a resonance detecting algorithm to each simulation in order to establish the distribution of mean-motion resonances between the remaining super-Earth mass planets in the final systems. Two particles $a$ and $b$ are considered to be in the $p : p + q$ mean-motion resonance if the resonant argument
\begin{equation}
\varphi = (p+q) \lambda_b - p \lambda_a - q \varpi_b
\label{eq:resarg}
\end{equation}
librates rather than circulates, where $\lambda = M + \varpi$ is the mean longitude, $M$ is the mean anomaly and $\varpi$ is the longitude of pericentre \citep{MurrayDermott}. We look for evidence of resonant behaviour in the last 20,000 yr (200 snapshots) of each simulation. At each of these snapshots we calculate the period ratio between each pair of adjacent planets, and then find the nearest rational number to this ratio in the form $p/(p+q)$. We impose the limit that $p, p+q < 9$\footnote{This limit on the magnitude of $p$ and $p+q$ is imposed to prevent spurious detection of very weak resonances.}. Equation \ref{eq:resarg} is then used to calculate the resonant argument across the last 200 snapshots with these values of $p$ and $q$. The algorithm looks for evidence of circulation in the sequence of resonant arguments. If the current sequence has a) a mean between $3\pi/4$ and $5\pi/4$, b) a range larger than $5.75$ and c) a standard deviation larger than $1.25$, it is considered to be circulating.  Once circulation is detected, then the algorithm begins building a new sequence from the point at which the last one ended, again looking for circulation. If no circulation is detected within at least the final 5000 yr of the simulation, the resonant argument is considered to be librating at the end of the simulation and thus the planets are in resonance. The definition of libration and circulation used by the algorithm is necessarily arbitrary and values for all the limits have been tweaked by hand to avoid false positives (due to the low sampling frequency of our output snapshots). In all cases the sampling frequency of the resonant argument is much lower than the orbital frequency of either planet, but examination of the evolution of several hundred resonant arguments by eye suggests that this algorithm returns very few false positives.

\section{Results}
\begin{table*}
\begin{tabular}{ l  c  c  c  c  c}
  \hline                      
   & \textbf{Planet e, $1 \mathrm{M_{jup}}$} & \textbf{Planet e,  $3 \mathrm{M_{jup}}$} &  \textbf{Planet f, $1 \mathrm{M_{jup}}$}& \textbf{Planet f,  $3 \mathrm{M_{jup}}$}
   & \textbf{Linear growth} \\
  \hline 
    \hline 
  \textbf{Planet b} & 2.16  $\times 10^{-4}$ & 1.21 $\times 10^{-4}$ & 5.85 $\times 10^{-4}$& 1.69 $\times 10^{-4}$& 1.80$\times 10^{-8}$\\
  \hline 
  \textbf{Planet c} & 3.03 $\times 10^{-5}$& 8.60 $\times 10^{-8}$ & 8.26 $\times 10^{-3}$& 3.12$\times 10^{-4}$&1.20$\times 10^{-8}$\\
  \hline  
  \textbf{Planet d} & 2.79$\times 10^{-8}$& 3.00 $\times 10^{-6}$ & 3.74 $\times 10^{-3}$& 2.27$\times 10^{-5}$&6.53$\times 10^{-5}$\\
  \hline  
  \textbf{Planet e} & N/A & N/A & 4.88 $\times 10^{-4}$& 1.96 $\times 10^{-6}$& 8.49$\times 10^{-9}$\\
  	 \hline  
  \textbf{Planet f} & N/A & N/A & N/A & N/A & 3.66 $\times 10^{-32}$\\
  \hline  

  \label{kstab}
\end{tabular}
\caption{Probabilities from the K-S test for each case with a giant planet in comparison to the control case. Each value represents the probability that the distribution of the position of each planet across all successful runs was picked from the same underlying distribution.}
\label{tab:kstable}
\end{table*}

\begin{table}
\begin{center}
\begin{tabular}{ r c c c c c c }
\hline
System& S & S/O & S/U & T & C & E \\ \hline \hline 
No giant (control) & 801 & 774 & 27 & 1 & 195 & 4\\ \hline
Planet e, $1 \mathrm{M_{jup}}$ & 760 & 726 & 34 & 0 & 234 & 6\\ \hline
Planet e, $3 \mathrm{M_{jup}}$ & 599 & 568 & 31 & 0 & 360 & 41\\ \hline
Planet f, $1 \mathrm{M_{jup}}$ & 772 & 731 & 41 & 0 & 223 & 5 \\ \hline
Planet f, $3 \mathrm{M_{jup}}$ & 705 & 660 & 45 & 1 & 279 & 16 \\ \hline
Linear growth                  & 912 & 912 & 0 & 1 & 87 & 1 \\ \hline

\end{tabular}
\caption{Outcome types for each set of 1000 runs. S: Runs that finished without a collision or ejection event. S/O: Subset of S that finished with no planets having swapped positions. S/U: Subset of S in which some planets swapped positions. T: Number of runs from S that were stopped once simulation time exceeded $15\tau$. C: Number of runs that ended in a collision between two planets. E: Number of runs that ended in a planetary ejection. \label{outcomes}}
\end{center}
\end{table}

The results of the K-S tests comparing the distribution of final planet positions between the control case and the cases with a giant are shown in table \ref{tab:kstable}. Figure \ref{fig:distdiff} shows exactly how these distributions compare to the control in cases where planet e or f becomes a giant, while figure \ref{fig:resdist} illustrates how the presence of a giant affects the final distribution of mean-motion resonances among the super-Earths. It is clear from the K-S test results that the presence of a giant planet during the evolution of compact systems can have a significant effect on their final structure. Figure \ref{fig:distdiff} suggests that the effect of a giant planet generally tends to push the interior super-Earths into more tightly packed orbits. This figure also illustrates how making the giant planet more massive amplifies this effect, with planets b and c being found on tighter orbits in the $3\mathrm{M_{Jup}}$ case than the $1\mathrm{M_{Jup}}$ case. The reason for this effect is evident from figure \ref{fig:resdist}. Compared to the control case, the cases with giant planets show a lower incidence of widely-spaced resonances (such as 2:1) with a correspondingly higher incidence of tighter resonances (such as 3:4). This suggests that the dynamical effect of the giant is to break the interior super-Earths out of wide resonances, allowing convergent migration to push them into tighter ones, which naturally results in more tightly-packed orbits. The linear growth model on the other hand appears to cause a dramatic increase in the occupancy of the 2:3 resonance, at the expense of both more- and less-tightly packed resonances. The dynamical reasonaing for this is clear: having the outer planets spend most of the simulation time at lower masses means they perturb the inner planets to a lesser extent, meaning that once the interior planets are in the 2:3 resonance, they are unlikely to break out.

It is worth considering how the incidence of resonances in our simulations compares to observations. With only one potential example of an observed system similar to those we form, this is naturally a difficult prospect. Nevertheless, \cite{Cabrera2014} note that the three super-Earths in Kepler-90 appear to be close to a 2:3:4 Laplace resonance. A cursory examination of our results shows that the incidence of this particular resonance chain more than doubles between the control case and the case with planet f becoming a $1 \mathrm{M_{jup}}$ planet, being present in $3.5\%$ of the successful runs in the former case and $8.2\%$ of runs in the latter case. A similar increase in the occurence of this resonance chain is seen in the case where planet e becomes a $1 \mathrm{M_{jup}}$ planet, with $5.9\%$ of runs exhibiting this behaviour. This suggests that outer giant companions preferentially lead to the formation of resonant chains between super-Earths.
 
The distribution of the planet neighbouring the giant is always significantly different than in the control case, but interior planets are affected to differing extents. For instance, it is clear from figure \ref{fig:distdiff} that the orbit of planet c is significantly altered by planet e becoming a giant, but in the case where planet f becomes a $1\mathrm{M_{Jup}}$ or $3\mathrm{M_{Jup}}$ giant, the distribution of planet c is altered to a lesser extent, with the difference to the control distribution becing more exagerrated in the former case. The giants are the same mass and at the same location in all of these cases, the only difference being the addition of an extra super-Earth (planet e) when planet f becomes a giant. This suggests that having an extra super-Earth between a planet and the giant can act to shield the planet from the dynamical effect of the giant. The size of this effect depends upon the final mass of the giant and the position of the perturbed planet relative to the giant, with larger giants having a more significant effect. 

The breakdown of simulation outcomes in table \ref{outcomes} reveals more about the effect the presence of a giant has on the evolution of the super-Earths. Larger giants naturally result in more collisions and ejections, suggesting that there may be a lower incidence of compact systems with very high mass outer planets, or at the very least a trend for fewer super-Earth mass planets in such systems. We note that the 4 ejections in the ``no-giant'' scenario are all caused by the innermost planet being scattered by its nearest neighbour into the path of planet c, which is almost 4 times larger than planet a. The encounter between planet a and planet c is then sufficient to push planet a onto a marginally hyperbolic orbit ($e \approx 1.01$). We also note that in a very small minority of cases, the giant planet does not complete its growth before the simulation ends. However, this is only a tiny fraction of our runs ($\approx 2\%$) and the planet is still generally many times larger than the super-Earths, so we do not count this as a separate outcome.

\section{Discussion}
\subsection{Implications}
Our results suggest that giant companions could affect compact systems in a similar way to disc turbulence; causing the breakdown of widely-spaced mean-motion resonances, and allowing the formation of tighter ones. This effect was explored in the context of disc turbulence by \cite{Rein2012}. We thus suggest that compact systems with tighter mean-motion resonances provide better candidates in searches for giant companions, since the tighter resonances may indicate that a giant has allowed the super-Earths to migrate through more widely-spaced ones. This mechanism may help to explain the formation of systems such as Kepler-36, the two (known) planets of which are near to the $7:6$ resonance \citep{Carter2012}. \cite{Paardekooper2013} suggested that this may be the result of turbulence in the disc breaking wide resonances and thus allowing convergent migration to push the 2 planets into closer orbits. We propose that an exterior giant companion could provide an alternate formation channel for such systems.

Similarly, it seems that the formation of the 2:3:4 Laplace resonance is amplified by the presence of a giant planet. This leads us to believe that systems of known super-Earths exhibiting such a resonant configuration would also be good candidates in follow-up searches for giant companions.

We also note that a more realistic approach may be some combination of our linear and runaway growth models, allowing all of the super-Earths to grow linearly to their final masses before allowing one of them to undergo runaway growth. Whilst the linear growth scenario cannot explain the formation of extremely tightly-packed systems such as Kepler-36, it does allow push a significant number of planetary pairs interior to the 2:1 resonance relative to the control case. The combination of this effect with the later pertubation caused by a giant could lead to even higher occupation of very tight resonances.

\subsection{Observability}
As we hypothesise that the giant planets in our simulations would not be seen by {\it Kepler}, we now estimate what fraction of these planets would actually be detectable. Since our simulations are all co-planar (by construction), we have to make some assumption regarding the inclination of the giant relative to the rest of the system in order to say whether or not the giant will transit. \cite{Fabrycky2014} found by comparing transit impact parameters of adjacent planets in compact systems that the mutual inclinations were in the range 1-2.2$^{\circ}$. We assume that this distribution extends to our giants also, and therefore assign random inclinations to our giants from a Gaussian distribution with a mean of 1.6$^{\circ}$ and a standard deviation of 0.6$^{\circ}$, such that the entire range suggested by \cite{Fabrycky2014} is included within $1 \sigma$ of the mean. The distribution is truncated at $0^{\circ}$ and $3.2^{\circ}$. We assume that the giant in each simulation transits if
\begin{equation}
\sin i \leq \frac{R_p + R_*}{a}
\end{equation}
where $i$ is the randomly assigned inclination, $R_p$ is the planetary radius, $R_*$ is the stellar radius (set to $R_\odot$) and $a$ is the semi-major axis of the planetary orbit. This condition ensures that the giant will be transiting regardless of the longitude of periapsis. Note that since we halt migration once the runaway growth phase begins, all of the giants in our models are at almost exactly 1AU, apart from small deviations caused by dynamical interactions with the other planets. Nevertheless, for the sake of self-consistency, we take the semi-major axes of the giants straight from our models, assign a random inclination, and then determine if they will transit using the above criterion. The results of this test are contained in table \ref{tab:transit}. It is clear that the vast majority ($\approx 99 \%$) of these giant companions would be undetectable via transit even with this rather modest inclination distribution. Note however that two factors could change the frequency of transits predicted by this model. Firstly, some of the planets excluded by this simple model as non-transiting would in fact be visible as transits assuming that their longitude of periapsis was such that they passed betweeen the star and the line of sight. Secondly, the radius $R_p$ assigned to the giants using the power-law from section \ref{sec:init} our $1 \mathrm{M}_{jup}$ models is approximately 1.5 times larger than the actual radius of Jupiter. Hence a small minority of planets that would nominally appear as grazing transits would not in fact be visible. In spite of these factors, we feel that these figures represent a good estimate of what percentage of giant companions would be visible to transit studies, and suggest that there could be a not-insignificant number of {\it Kepler} systems harbouring unseen giants.
\begin{table}
\begin{center}
\begin{tabular}{ c c c c  }
\hline
  \textbf{Planet e, $1 \mathrm{M_{jup}}$} & \textbf{Planet e,  $3 \mathrm{M_{jup}}$} &  \textbf{Planet f, $1 \mathrm{M_{jup}}$}& \textbf{Planet f,  $3 \mathrm{M_{jup}}$} \\ \hline \hline
0.96\%  & 0.88\% & 1.23\% & 0.91\%\\ \hline

\end{tabular}
\caption{Percentage of giants found to be detectable by transit using our simplified transit model. \label{tab:transit}}
\end{center}
\end{table}	
\begin{figure}
\begin{center}
\includegraphics[width=0.99\linewidth]{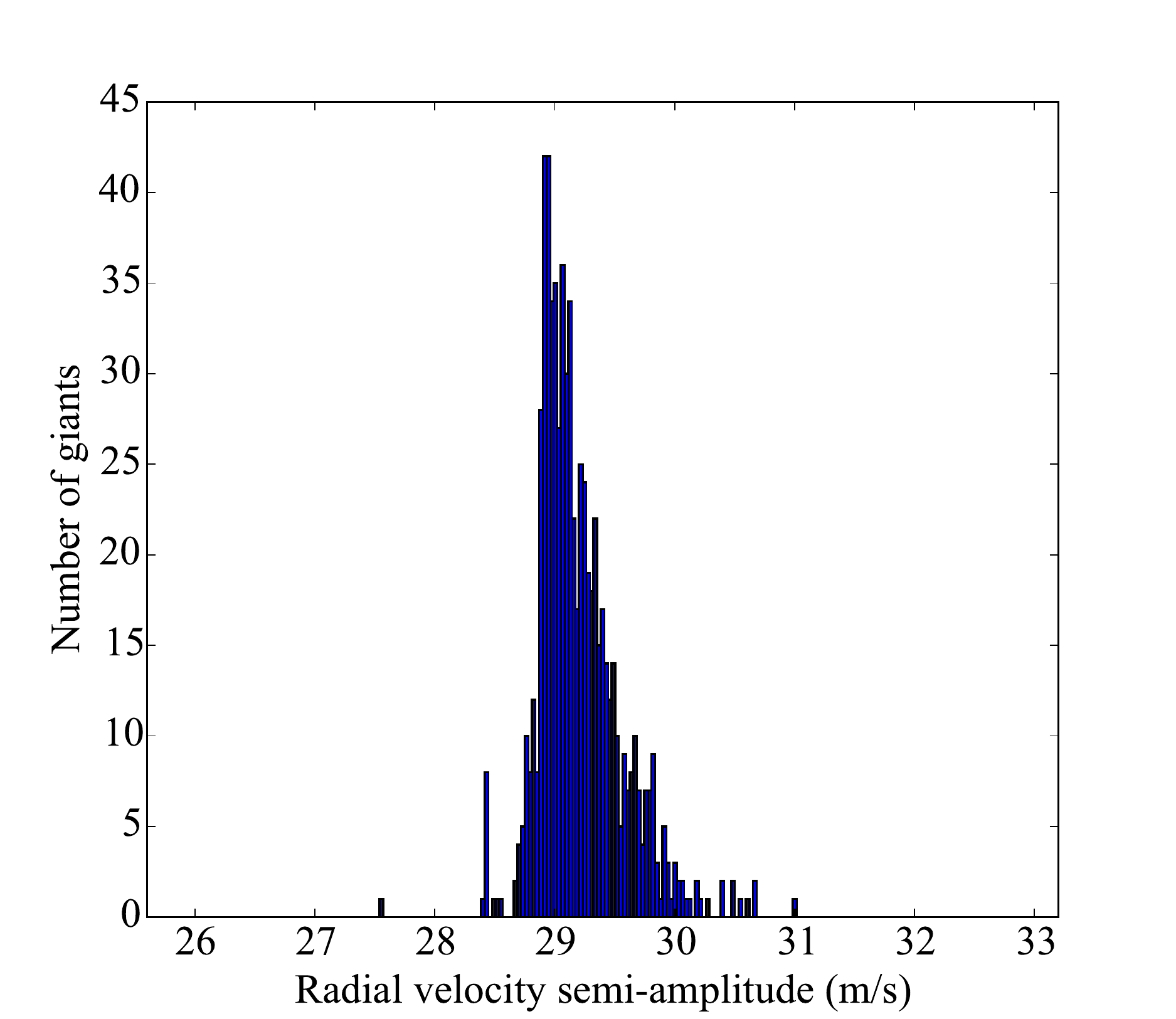}
\caption{Example radial velocity distribution for our models where planet e becomes a $1 \mathrm{M_{jup}}$ giant.\label{fig:rv} }
  \end{center}
\end{figure}

Using the same inclination distribution, we can also establish if such giants would be detectable by radial velocity (RV) surveys. We use equation 1 from \cite{Fischer2014} to calculate radial velocities for a planets. An example distribution can be seen in figure \ref{fig:rv}: the low range of inclinations gives only a small spread in reflex velocities, and typical signals are $\gtrsim25$m/s. For bright stars such a signal is easily detectable via RV observations \citep[see e.g.,][]{Mayor2011}. However, the majority of {\it Kepler} host stars are much too faint for RV follow-up, and detection would be further hindered by the relatively long time-scales ($\gtrsim1$yr) on which the RV signal oscillates. With a dedicated campaign it may be possible to detect ``hidden giants'' around the brightest {\it Kepler} host stars, but otherwise they are likely to remain undetected by the current generation of planet-hunting facilities.

\subsection{Limitations}
There are a number of necessary limitations imposed upon these models in order to reduce the vast parameter space to something computationally-viable. The majority of these, such as the arbitrary stopping criterion and an over-prediction of the abundance of MMRs, are discussed by \cite{Hands2014}. However, the introduction of the runaway growth prescription brings about several more parameters worth consideration. The growth time-scale $\tau_g$  and final mass of a planet undergoing runaway growth depend sensitively upon the structure and composition of the disc and will naturally vary from case-to-case in reality. A full exploration of the effect that changing these parameters has is beyond the scope of this study, and we believe $10^3\mathrm{yr}$ to be a reasonable estimate of the timescale at the small orbital radii considered in this proof-of-concept study.  The choice of radius at which runaway growth begins (1AU) is also arbitrary, and in this case was chosen to be the minimum radius at which a giant planet might be expected to exist without being detected by {\it Kepler} as a transiting planet. Thus the pertubations caused by the giants in our models are the maximum effect that one might expect to see in a tightly-packed {\it Kepler} system, and any real giants might have a less significant impact. Future work could concentrate on how the degree of pertubation changes as the spacing between the giant and the star is changed.

\section{Summary}
In this paper we have investigated the dynamical impact of a giant companion on the formation of tightly-packed planetary systems. A giant planet can break widely-spaced mean-motion resonances and push compact systems into tighter ones, leading to more tightly-packed orbits and to the formation of tight, Laplace-resonant chains. The magnitude of this effect is dependent upon which of the planets becomes a giant, with planets that are nearer to the giant being more strongly perturbed, and also increases for more massive giant planets. We suggest that this could provide an alternate channel for the assembly of {\it Kepler} systems that are close to tight resonances, and that in turn these systems may prove to be promising candidates in searches for far-out giant companions.

\section*{Acknowledgements}
The authors would like to thank Hossam Aly and Walter Dehnen for useful discussions, and the anonymous reviewer for comments which greatly improved the manuscript. TOH is supported by an Science \& Technology Facilities Council (STFC) PhD studentship. RDA acknowledges support from STFC through an Advanced Fellowship (ST/G00711X/1), and from the Leverhulme Trust through a Philip Leverhulme Prize. Astrophysical research at the University of Leicester is supported by an STFC Consolidated Grant (ST/K001000/1). This research used the ALICE High Performance Computing Facility at the University of Leicester. Some resources on ALICE form part of the DiRAC Facility jointly funded by STFC and the Large Facilities Capital Fund of BIS.  This work also used the DiRAC {\it Complexity} system, operated by the University of Leicester IT Services, which forms part of the STFC DiRAC HPC Facility ({\tt http://www.dirac.ac.uk}). This equipment is funded by BIS National E-Infrastructure capital grant ST/K000373/1 and  STFC DiRAC Operations grant ST/K0003259/1. DiRAC is part of the UK National E-Infrastructure.

\bibliographystyle{mnras}
\bibliography{packed}
\appendix

\label{lastpage}

\end{document}